\documentclass[aps,prb,twocolumn,superscriptaddress,floatfix]{revtex4}

\usepackage{amscd,rotating}
\usepackage{amsmath} 
\usepackage{amssymb}
\usepackage{bm}
\usepackage{graphicx}     
\usepackage{longtable}
\usepackage{color}

\begin{document}


\title{Effective action for phase fluctuations in $d$-wave superconductors
near a Mott transition}

\author{Damien Bensimon}
\affiliation{Max-Planck-Institut f\"ur Festk\"orperforschung,  
Heisenbergstrasse 1, D-70569 Stuttgart, Germany.}

\author{Roland Zeyher}
\affiliation{Max-Planck-Institut f\"ur Festk\"orperforschung,  
Heisenbergstrasse 1, D-70569 Stuttgart, Germany.}

\date{\today}


\begin{abstract}
Phase fluctuations of a $d$-wave superconducting order parameter are 
theoretically studied in the context of high-T$_c$ cuprates. We consider 
the $t-J$ model describing layered compounds, where the Heisenberg interaction 
is decoupled by a $d$-wave order parameter in the particle-particle channel.
Assuming first that the 
equilibirum state has long-range phase order, the 
effective action $\mathcal{S}_{eff}$ is derived perturbatively for small 
fluctuations within a path integral formalism, in the presence of the 
Coulomb and Hubbard interaction terms. In a second step, a more general 
derivation of $\mathcal{S}_{eff}$ is performed in terms of a gradient 
expansion which only assumes that the gradients of the order parameter are 
small whereas the value of the phase may be large. We show that in the
phase-only approximation the resulting $\mathcal{S}_{eff}$ reduces in 
leading order in the field gradients to
the perturbative one which thus allows to treat also the case without 
long-range phase order or vortices. Our result generalizes previous
expressions for $\mathcal{S}_{eff}$ to the case of interacting electrons,
is explicitly gauge invariant, and avoids problematic singular
gauge transformations.  

\end{abstract} 

\maketitle


\section{Introduction}
\label{Sec-Intro}

The nature of the pseudogap in underdoped high-T$_c$ oxides and its
relation to superconductivity remains one of the unsolved problems in 
high-T$_c$ superconductivity. One scenario assumes that the physics 
in the underdoped and optimally doped region is mainly determined
by the competition of the superconducting order parameter with a second
one in the particle-hole channel. Possible candidates are antiferromagnetism,
charge- and spin-density waves,\cite{Castro2001,Prelovsek2002} in particular 
with an internal $d$-wave symmetry, \cite{Cappelluti1999,Chakravarty2001}
and stripes.\cite{Tranquada1995} One experimental constraint 
is that this second
order parameter is strongly anisotropic, being large along the $k_x$
and $k_y$ axis of the Brillouin zone,
and practically zero near the diagonal $k_x=k_y$. This requirement
is most naturally fulfilled if this order parameter has $d$-wave
symmetry like the superconducting order parameter.   
A second scenario assumes that no instabilities
or strong fluctuations in the particle-hole channel are relevant in the
underdoped regime but that phase (and perhaps amplitude) fluctuations
of the superconducting order parameter are important in this region.
\cite{Emery}
  
Presently, it is unclear which of the above two scenarios is more
realistic in describing high-T$_c$ cuprates. In this paper we want
to explore some aspects of the second approach in more detail.
While it seems that the first scenario has become more popular than the 
second one we think that there are enough reasons to
study phase fluctions of the superconducting order parameter.
For instance, order parameters due to structural phase transitions
with finite momentum transfers may easily be anisotropic because of
strongly varying nesting properties along the Fermi surface. However,
whether one can achieve along this route the observed large anisotropy
or even d-wave symmetry of the pseudogap in a generic way remains unclear.
Experimental SIN tunneling data in the cuprates also show a rather symmetric 
one-particle density of states, both in the superconducting and the
pseudogap region, with respect to the chemical potential.\cite{Miyakawa1999}
Such a behavior
is characteristic for superconductivity but not generic for
densities with structural order parameters in the particle-hole channel, 
especially, if they are associated with large momentum transfers. 
These problems vanish if one assumes that only one complex order parameter 
related to the observed $d$-wave superconductivity is relevant. 

Fluctuations effects in superconductors are often described by
the time-dependent Ginzburg-Landau equations.\cite{Larkin2005} They describe a
relaxational behavior of the order parameter back to equilibrium,
i.e., their time-dependent part is of first order in the time
derivative and independent of momentum. In the pure system they
are most successfull near T$_c$ and at low frequencies compared
to the temperature. In the following we want to consider a different
regime, namely, temperatures substantially below the mean-field
transition to superconductivity where the mean-field order parameter 
is large but varies slowly in space and time because of 
a small phase stiffness constant $\Lambda$.\cite{Uemura} The $t$-$J$ model
and its constraint to have no doubly occupied sites leads necessarily
to small values for $\Lambda$'s at small dopings. At low temperature
$\Lambda$ is determined by the diamagnetic term  which is given by the
Fourier transform of the one-particle momentum distribution function at 
non-vanishing lattice vectors. The constraint of having no doubly occupied 
sites implies that the maximum occupation of a momentum state is $1+\delta$ 
instead of $2$ as in the free case.\cite{Wang-1992} 
This together with the sum rule shows 
that the diamagnetic term must vanish linearly in $\delta$ in the limit 
$\delta \rightarrow 0$. As a result large phase fluctuations should occur 
at small dopings. 
In agreement with this picture we derive in the present work an effective
action for phase fluctuations not as a power but as a gradient expansion in 
the phase. The method we follow also allows to take into account 
interaction terms
between the electrons such as the Hubbard or the long-range Coulomb
interaction. This feature is important because the small phase stiffness
at low dopings is caused by correlation effects. At the same time our
treatment leads to an explicitly gauge invariant form for the effective
action for phase fluctuations in contrast to many previous treatments. 

Most of the previous derivations of the effective action for phase 
fluctuations employed singular gauge 
transformations\cite{Ambeg-1984,Param-PRB2000,Carbotte2004,Benfatto}. 
Such a transformation means in the static case for the
Bogoliubov equations that 
$\Delta(\bm{r}) \rightarrow
\Delta(\bm{r})\cdot e^{-\mathrm{i}\Phi(\bm{r})}$, 
$u(\bm{r}) \rightarrow u(\bm{r})\cdot e^{-\mathrm{i}\Phi(\bm{r})/2}$,
$v(\bm{r}) \rightarrow v(\bm{r})\cdot e^{\mathrm{i}\Phi(\bm{r})/2}$,
$\bm{A}(\bm{r}) \rightarrow \bm{A}(\bm{r}) - (\hbar c/2\mathrm{i}e)\nabla 
\Phi(\bm{r})$. \cite{Ketterson1999}
$\Delta,u,v$ are the pair, electron and hole wave functions,
respectively, ${\bf A}$ the vector potential, and $\Phi$ is an
arbitrary function. This transformation
allows to remove completely the fluctuations in $\Delta$ and to describe them
as fluctuations in ${\bf A}$. 
However, $\Delta$, $u$ and $v$ have to be unique functions
of $\bm{r}$, i.e., their phase can only change by multiples of $2\pi$ 
after passing through a closed loop. This condition is in general violated
in the presence of vortices after having performed such 
a gauge transformation. The importance of a possible
non-uniqueness of wave functions for physical quantities \cite{Anders-cm1998}
is presently not clear, most authors 
neglect this problem whereas, according to Refs. 
\onlinecite{FrTesa-PRL2001,FrTesa-PRB2002,VaTesa-PRL2003}, 
it causes very interesting effects such as non-Fermi liquid behavior, 
power laws of correlation functions, etc., in the normal state. 
In order not to violate the basic requirement of uniqueness of wave functions
we avoid completely problematic gauge transformations and
derive the effective action for phase
fluctuations by means of a gradient expansion in the order parameter.

The paper is organized as follows. In section II the Hamiltonian
is specified and phase fluctuations of the order parameter
are introduced. In section III the effective action for phase fluctuations
is derived under the assumption that the phases $\theta$ deviate only little
from a homogenous state, for instance, $\theta = 0$. 
The microscopic quantities appearing in the action are one- and two-particle
Green's functions associated with the Hamiltonian without external
potentials or phase fluctuations.
A more general derivation for the effective action is given in section IV 
using a gradient expansion which
only assumes that the gradients of the order parameter are small whereas
$\theta$ may be large. It is shown that the two effective actions
derived in sections III and IV are equivalent in the phase-only 
approximation for the order parameter. Section V contains a discussion
of the results and the conclusions. In section III relations between
different correlation functions describing density, current and pair
fluctuations have been inferred from non-singular gauge transformations
and used in obtaining the final form for the effective action. 
We found it useful to check these relations in the non-interacting case
and in the long-wavelength, low-frequency limit directly without 
referring to gauge invariance arguments. The details of this calculation 
are given 
in the appendix.



\section{Hamiltonian and phase fluctuations of the order parameter}
\label{Sec-Model-MFT}

We consider a generalized $t$-$J$ model,\cite{Anderson1987,Rice1988}
which also contains repulsive Coulomb and Hubbard interactions, on a  
Bravais lattice consisting of layers of squares along the $x$ and $y$
axis. Its Hamiltonian is given by 
\begin{eqnarray}
\mathcal{H} &=& - \sum_{i,j,\sigma} t_{ij} 
c_{i\sigma}^{\dagger}c_{j\sigma}  
+ J \sum_{\langle i,j \rangle } \bigg\{ {\mathbf{S}}_{i} \cdot 
{\mathbf{S}}_{j}  - \frac{1}{4} n_{i} n_{j} \bigg\} \nonumber \\
& &+\mathcal{H}_{int}~,
\label{HtJ-def} \\
\mathcal{H}_{int} &=& \frac{1}{2} \sum_{i,j} V_{ij} n_{i} n_{j} + \frac{U}{2} 
\sum_{i} n_{i} n_{i}~.
\label{Hint}
\end{eqnarray}
$\sigma$ is the SU(2) spin color, $c_{i\sigma}^{\dagger}$ ($c_{i\sigma}$) 
the creation 
(annihilation) operator of a spin $\sigma$ electron on the site $i$, $J$ the 
Heisenberg 
interaction, $V_{ij}$ the Coulomb interaction between the sites $i$ and $j$, 
and 
$U$ a repulsive Hubbard term. 
$t_{ij}$ is the electronic hopping term between the sites $i$ and $j$, 
$\langle ~ \rangle$  denotes a pair index for nearest neighbor sites
on the same layer. 
${\mathbf{S}}_{i}$ and $n_{i}$ are 
the spin and occupancy number operators of site $i$, respectively. 
In order to simplify the notation later we will put $a= \hbar =c =1$,
where $a$ is the lattice constant of the square lattice and $c$ the 
velocity of light. We also assume that the lattice contains  $\mathrm{N_s}$
sites. Let us   
introduce the singlet pair operators
\begin{eqnarray}
B_{i j}^{\dagger} = c_{i \uparrow}^{\dagger} c_{j \downarrow}^{\dagger} - 
c_{i \downarrow}^{\dagger} c_{j \uparrow}^{\dagger} ~,~
B_{i j} = c_{j \downarrow} c_{i \uparrow} - c_{j \uparrow} c_{i \downarrow} ~,
\label{BondOp-def}
\end{eqnarray}
which respectively create and annihilate a singlet on the bond 
$\langle i,j \rangle$. $\mathcal{H}$ can then 
be expressed equivalently as \cite{Param-PRB2000}
\begin{equation}
\mathcal{H} = -\sum_{i,j,\sigma} t_{ij}
c_{i \sigma}^{\dagger}c_{j \sigma} 
- \frac{J}{2} \sum_{\langle i,j \rangle} B_{i j}^{\dagger} B_{i j} 
+\mathcal{H}_{int}~.
\label{HtJ-BondOp}
\end{equation}

To derive the partition function, we work in a path integral formalism. 
\cite{NegOr,Tsvelik-book,Nagaosa_bookCM,Nagaosa_bookSC} The Heisenberg 
interaction term is decoupled using the Hubbard-Stratonovitch 
transformation,\cite{Strato,Hub} the resulting  
complex fields $\Delta^{*}$, $\Delta$ correspond to the 
superconducting order parameter. \cite{Nagaosa_ref} The partition function 
of the $t-J$ model (\ref{HtJ-BondOp}) in the imaginary time formalism
is given by
\begin{eqnarray}
 {\mathcal{Z}} = \int \mathcal{D}\bar{\Psi} \mathcal{D}\Psi \mathcal{D}\Delta 
\mathcal{D}\Delta^{*} ~
 \mathrm{exp} \big(- {\mathcal{S}} \big)~,
 \label{ZtJ-def}
\end{eqnarray}
with the action
\begin{eqnarray}
{\mathcal{S}} &=& \int_{0}^{\beta}d\tau \Bigg[
\sum_{i,\sigma}  \bar{\Psi}_{i \sigma}(\tau) \big\{ \partial_{\tau} - 
\mu\big\} \Psi_{i \sigma}(\tau) 
\nonumber \\ 
& &\hspace{11mm}
-~ \sum_{i,j,\sigma} t_{ij} 
\bar{\Psi}_{i \sigma}(\tau) \Psi_{j \sigma}(\tau)  
\nonumber \\ 
 & & \hspace{11mm} 
 -~ \sum_{\langle i,j \rangle} \bigg\{ 
 \frac{1}{4} \big[ \Delta_{ij}(\tau) \bar{B}_{ij}^{*}(\tau) +  
\Delta_{ij}^{*}(\tau) \bar{B}_{ij}(\tau) \big]
\nonumber \\ 
& &\hspace{23mm}
-~  \frac{1}{8J} \mid 
\Delta_{ij}(\tau) \mid^{2} \bigg\} \Bigg] 
+ \mathcal{S}_{int}~.
\label{StJ-def}
\end{eqnarray}
$\mathcal{S}_{int}$ is the contribution to the action due to 
$\mathcal{H}_{int}$, $\beta$ is the thermal factor,
$\beta = 1/k_{\mathrm{B}}T$,
$\mu$ is the chemical potential controlling the electron density while 
$\bar{\Psi}$, $\Psi$ and $\bar{B}^{*}$, $\bar{B}$ are Grassmann variables 
corresponding to the electronic operators $c^{\dagger}$, $c$ and singlet pair 
operators $B^{\dagger}$, $B$, respectively.

The complex bond variables $\Delta_{i,i+\hat{x}}$ and $\Delta_{i,i+\hat{y}}$ 
may be written without loss of generality as 
\begin{eqnarray}
\label{Delta-gen}
& &\Delta_{ij}(\tau) = \mid \Delta_{ij}(\tau) \mid \gamma_{ij} 
\mathrm{e}^{\mathrm{i}\phi_{ij}(\tau)} 
~,~ \\
& &\gamma_{ij} = \left\{ \begin{array}{ll} +1 & \textrm{for~} j = i+\hat{x} \\
                                        -1 & \textrm{for~} j = i+\hat{y} \end{array} 
\right.~, \nonumber
\end{eqnarray}
where $\hat{x}$ and $\hat{y}$ are basis vectors of the direct lattice. 
The equilibrium $d$-wave order parameter value is given by
\begin{eqnarray}
\mid \Delta_{ij}(\tau) \mid \equiv \Delta_{0}~,~\phi_{ij}(\tau) \equiv 0~,
\label{Delta-dwaveEq}
\end{eqnarray}
where $\Delta_{0}$ is a real number independent of $i$ and $\tau$.
In the present work, the fluctuations of the amplitude are neglected. 
The spatial variations of the phase $\phi$ are supposed to be 
independent of the $x$ and $y$ 
directions. This allows us to introduce the average phase of the two 
bonds, and to put the phase variable on the lattice points, i.e. 
\cite{Param-PRB2000}
\begin{eqnarray}
\phi_{ij}(\tau) = \frac{1}{2} \big[ \theta_{i}(\tau) 
+ \theta_{j}(\tau)\big]~,
\label{Phi-def}
\end{eqnarray}
where $\theta_{i}$ is a real field. The fluctuating 
order parameter is thus assumed to be
 \begin{eqnarray}
  \label{DeltaPhas-def}
 \Delta_{i,i+\hat{x}}(\tau) = \Delta_{0} \cdot
 \mathrm{e}^{\mathrm{i} [ \theta_{i}(\tau) + 
\theta_{i+\hat{x}}(\tau)]/2} ~,~ \\
 \Delta_{i,i+\hat{y}}(\tau) = -\Delta_{0} \cdot
 \mathrm{e}^{\mathrm{i} [ \theta_{i}(\tau) + 
\theta_{i+\hat{y}}(\tau)]/2}~.
\nonumber
\end{eqnarray}

It is well-known that phase fluctuations induce charge and current 
fluctuations and thus electric and magnetic 
fields. \cite{DeGennes-book} Therefore the resulting electromagnetic field 
has to be considered in a general formalism by including 
scalar and vector potentials in the action, especially, to guarantee 
gauge invariance. In the case of lattice models, a minimum coupling 
scheme to describe the interactions involving the electromagnetic 
field is given by the Peierls substitution, \cite{Peierls-1933} 
which corresponds to
 \begin{eqnarray}
- \mu \cdot \bar{\Psi}_{i \sigma}(\tau) \Psi_{i \sigma}(\tau)
&\longrightarrow& \Big[ -\mu - e \cdot A_{0}({\bm{r}}_{i},\tau) \Big]
\nonumber \\
& & \times~ \bar{\Psi}_{i \sigma}(\tau) \Psi_{i \sigma}(\tau) ~, 
\label{Peierls-def1} \\
t_{ij} \cdot  \bar{\Psi}_{i \sigma}(\tau) \Psi_{j \sigma}(\tau) 
&\longrightarrow& t_{ij} \cdot \mathrm{exp} \bigg(- \mathrm{i} e 
\int_{{\bm{r}}_{j}}^{{\bm{r}}_{i}}       
{\bm{A}}({\bm{r}},\tau)\cdot d{\bm{l}} 
\bigg) 
\nonumber \\
& & \times~ \bar{\Psi}_{i \sigma}(\tau) \Psi_{j \sigma}(\tau) ~,
 \label{Peierls-def2}
\end{eqnarray}
where $A_{0}$ and ${\bm{A}} \equiv (A_{x}, A_{y}, 
A_{z})$ are the scalar and vector potentials, respectively, and ($-e$) 
is the electron charge. In the case of slowly varying phase fluctuations
the induced electromagnetic potentials 
also vary slowly in time and space so that we can write \cite{GraVog-1995}
\begin{equation}
 \int_{{\bm{r}}_{j}}^{{\bm{r}}_{i}}       
{\bm{A}}({\bm{r}},\tau) \cdot d{\bm{l}}
 \approx ({\bm{r}}_{i} - {\bm{r}}_{j}) \cdot \frac{1}{2}
 \Big\{       {\bm{A}}({\bm{r}}_{j}, \tau) +        
{\bm{A}}({\bm{r}}_{i}, \tau) 
\Big\}~. \nonumber
\label{Peierls-tsimpl}
\end{equation}
 

\section{Effective action for small phase fluctuations}
\label{Sub-DerivEffAc}

In this section we first give explicit expressions for the change of the 
action away from the $d$-wave saddle point  
due to slow spatial and temporal phase fluctuations and the corresponding 
induced electromagnetic field. The effective action $\mathcal{S}_{e\!f\!f}$ 
is defined by 
\begin{eqnarray}
 \frac{{\mathcal{Z}} [A,\theta]}{{\mathcal{Z}}^{(0)}}
 = \mathrm{exp} \Big(- {\mathcal{S}}_{e\!f\!f} [A,\theta] \Big)~,
\label{SPhasEff-def}
\end{eqnarray}
using the abbreviation $A=(A_{0},{\bm{A}})$.
${\mathcal{Z}} [A,\theta]$ is the total partition function, containing also the
effects due to fluctuations in $A$ and $\theta$. 
${\mathcal{Z}}^{(0)}$ denotes the $d$-wave saddle point partition 
function in the presence of the Coulomb and Hubbard terms, and 
${\mathcal{S}}^{(0)}$ its corresponding action
\begin{eqnarray}
\label{Zdwave-def}
& &{\mathcal{Z}}^{(0)} = \int \mathcal{D}\bar{\Psi} \mathcal{D}\Psi \
\mathrm{exp} \Big(- {\mathcal{S}}^{(0)} \Big)~, \\
{\mathcal{S}}^{(0)} &=& \int_{0}^{\beta}d\tau \Bigg[
\sum_{i,\sigma} \bar{\Psi}_{i \sigma}(\tau) \big\{ \partial_{\tau} 
- \mu\big\} \Psi_{i \sigma}(\tau) 
\nonumber \\ 
& & \hspace{12mm}
-~ \sum_{i,j,\sigma} t_{ij} 
\bar{\Psi}_{i \sigma}(\tau) \Psi_{j \sigma}(\tau) \nonumber \\
& & \hspace{12mm} -~ \sum_{\langle i,j \rangle} \bigg\{ 
\frac{1}{4} \Delta_{0} \gamma_{ij} 
\big[  \bar{B}_{ij}^{*}(\tau) +  \bar{B}_{ij}(\tau) \big]
\nonumber \\ 
& & \hspace{25mm}
-~ \frac{1}{8J}  \big( \Delta_{0} \big)^{2} \bigg\} \Bigg] 
+ \mathcal{S}_{int}~. 
\label{Sdwave-def}
\end{eqnarray}

 By expanding perturbatively the  action (\ref{StJ-def}) with respect 
to the phase exponential factors and electromagnetic potentials appearing 
in Eqs.\! (\ref{DeltaPhas-def}) and (\ref{Peierls-def1}), 
(\ref{Peierls-def2}), 
respectively, one obtains the partition function and action related to 
fluctuations in the phase and the vector potential. We go 
over to imaginary frequencies and momentum space. The fermionic Matsubara 
frequencies 
are labelled by $\mathrm{i}\nu_{m}$, and
the bosonic ones by $\mathrm{i}\omega_{n}$. It is convenient 
to define the four-dimensional wave vectors 
$k=(k_\alpha)_{0 \leq \alpha \leq 3}=(\mathrm{i}\nu_m,\bm{k})$ 
and $q =(q_\alpha)_{0 \leq \alpha \leq 3}=(\mathrm{i}\omega_n,\bm{q})$.
The Fourier transformed field variables are defined by
\begin{eqnarray}
\label{Fourier}
\bar{\Psi}_\sigma(\mathrm{i}\nu_m,\bm{k}) &\equiv& \bar{\Psi}_\sigma(k) 
\\
&=& \frac{1}{\sqrt{\beta \mathrm{N_s}}}
\sum_i \int_0^\beta d \tau \mathrm{e}^{-\mathrm{i}\nu_m \tau + 
\mathrm{i}\bm{k} \cdot \bm{r}_i} 
\bar{\Psi}_{i\sigma}(\tau)~. \nonumber
\end{eqnarray}
It is also convenient to introduce 
Nambu spinor field operators \cite{Nambu-1960,Rick-GreenF} by
\begin{eqnarray}
\bar{\Phi}(k)
= \left( \bar{\Psi}_\uparrow(k) \ \ \
         \Psi_\downarrow(-k) \right) ~,~
\Phi(k) 
= \left( \begin{array}{c}  \Psi_\uparrow(k) \\
                           \bar{\Psi}_\downarrow(-k)
         \end{array} \right)~.~
\label{NambSpinor-def}
\end{eqnarray}
In this section we consider only terms up to the second order in
the fluctuations $\theta$. This means that we allow only small
phase fluctuations around a homogenous state characterized by $\theta = 0$.
The partition function becomes then
\begin{eqnarray}
& & \label{ZPhas-def}
{\mathcal{Z}} [A,\theta] = \int \mathcal{D}\bar{\Psi} \mathcal{D}\Psi \
\mathrm{exp} \Big(- {\mathcal{S}} [A,\theta] \Big)~, \\
 \label{SPhas-def}
& &{\mathcal{S}} [A,\theta] =  {\mathcal{S}}^{(0)} 
+ \mathcal{S}'_A + \mathcal{S}'_\theta~,\\
& &\mathcal{S}'_A =
\frac{1}{\sqrt{\beta \mathrm{N_s}}} \sum_{k,q}  \sum_{\alpha=0}^3 
v_\alpha (\bm{k})
A_\alpha(q) \tilde{\Phi}(n_\alpha,k,q)  \nonumber \\
& & \hspace{10mm} +~ \frac{1}{\beta \mathrm{N_s}}  \sum_{k,q,q'} 
\sum_{\alpha,\alpha'=1}^3 m_{\alpha \alpha'}^{-1}(\bm{k})
A_\alpha(q) A_{\alpha'}(q') \nonumber \\
& & \hspace{35mm} \times~ \tilde{\Phi}(3,k,q+q')~,  
\label{LA} \\
& & \mathcal{S}'_\theta  = \frac{1}{\sqrt{\beta \mathrm{N_s}}} 
\sum_{k,q} w(\bm{k},\bm{q}) 
\theta(q) \tilde{\Phi}(2,k,q)  \nonumber\\
& & \hspace{10mm}  +~ \frac{1}{\beta \mathrm{N_s}} \sum_{k,q,q'} 
z(\bm{k},\bm{q},\bm{q}') \theta(q) \theta(q')
\nonumber \\
& & \hspace{25mm} \times~ \tilde{\Phi}(1,k,q+q')~,
\label{LB}
\end{eqnarray}
with
\begin{eqnarray}
& & v_0(\bm{k}) = -e~,~
v_\alpha(\bm{k}) = (-e) \frac{\partial \epsilon_{\bm{k}}}{\partial k_\alpha} 
~~\mbox{for $\alpha = 1, 2, 3$}~, ~~~~~~~\\
& & m_{\alpha \alpha'}^{-1}(\bm{k}) = \frac{{e}^2}{2}
\frac{{\partial}^2 \epsilon_{\bm{k}}}
{\partial k_\alpha \partial k_{\alpha'}} ~~\mbox{for $\alpha, \alpha' = 
1, 2, 3$}~, \\
& & \epsilon_{\bm{k}} = - \sum_{\bm{r}_i - \bm{r}_j} t_{ij} 
\mathrm{e}^{\mathrm{i}\bm{k}\cdot(\bm{r}_i - \bm{r}_j)} ~,\\
& &w(\bm{k},\bm{q}) = \frac{1}{2}(\Delta_{\bm{k}} + \Delta_{\bm{k}+\bm{q}})~, 
\\
& &z(\bm{k},\bm{q},\bm{q}') = \frac{1}{8} 
(\Delta_{\bm{k}} + 2\Delta_{\bm{k}+\bm{q}}+ \Delta_{\bm{k}+\bm{q}+\bm{q}'})~, 
\\
& & \Delta_{\bm{k}} = \frac{\displaystyle \Delta_{0}}{\displaystyle 2} 
\big[\cos({k}_{x})-\cos({k}_{y})\big]~, \\
& &\tilde{\Phi}(\alpha,k,q) = \bar{\Phi}(k+ q)
{\bm{\sigma}}_{\alpha} \Phi(k)~.
\end{eqnarray}
$\epsilon_{\bm{k}}$ is the hopping energy and $\Delta_{\bm{k}}$ the 
$d$-wave superconducting order parameter.
${\bm{\sigma}}_{\alpha}$ denotes for $\alpha=1,2,3$
the Pauli matrices along the $x$, $y$, and $z$ directions, respectively,
and for $\alpha=0$ the $2 \times 2$ identity matrix. $n_\alpha$ is equal to 3 
for $\alpha=0$ and zero otherwise.  
The summations in $\bm{k}$-space always extend over the first Brillouin zone.

The next step consists in performing the functional integration over the 
$\bar{\Psi}$, $\Psi$ fermionic fields appearing in Eq. (\ref{ZPhas-def}). 
It is achieved by noticing that $\mathcal{Z} [A,\theta]/{\mathcal{Z}}^{(0)}$ 
can be seen as a generating functional, $\theta$, $A$ representing external 
sources which couple to one-particle density operators 
$\tilde{\Phi}(\alpha,k,q)$. Abbreviating the set of variables 
$\{\alpha_1,k_1,q_1\}$, $\{\alpha_2,k_2,q_2\}$ symbolically by $1$, $2$, 
etc., we can write
\begin{equation}
\mathcal{S}'_A + \mathcal{S}'_\theta = \int d 1 E(1) \tilde{\Phi}(1)~,
\end{equation}
where an explicit expression for $E(1)$ can easily be read off from 
Eqs.\! (\ref{LA}) and(\ref{LB}).
The integration over Fermi fields yields then for $\mathcal{S}_{e\!f\!f}$,
defined in Eq.\! (\ref{SPhasEff-def}),\cite{NegOr}
\begin{eqnarray}
\mathcal{S}_{e\!f\!f}[A,\theta] &=& \int d 1 ~\mathcal{G}_c(1) E(1) 
\nonumber \\
& &-~ \frac{1}{2} \int d 1 d 2 ~\mathcal{G}_c(1;2) E(1) E(2) + ...~,~~~
\label{Seff-NegOr}
\end{eqnarray} 
$\mathcal{G}_c$ are connected Green's functions. They are related 
to the usual Green's functions $\mathcal{G}$, defined by
\begin{equation}
\mathcal{G}(1;...;n) = \int \mathcal{D} \bar{\Phi} 
\mathcal{D} \Phi ~\tilde{\Phi}(1)...\tilde{\Phi}(n)
\mathrm{e}^{-\mathcal{S}_0}/\mathcal{Z}_0~,
\label{Green}
\end{equation}
via the cumulant expansion of $\mathcal{G}_c$ in terms of $\mathcal{G}$ 
obtained from the identity \cite{NegOr} 
\begin{equation}
\mathcal{W}[A,\theta] =
\mathrm{ln} \bigg(\frac{\mathcal{Z}[A,\theta]}{{\mathcal{Z}}^{(0)}}\bigg)~,
\label{Gener-connect}
\end{equation}
where $\mathcal{W}[A,\theta]$ is the generating functional for
connected Green's functions.
Explicitly, one obtains $\mathcal{G}_c(1) = 
\mathcal{G}(1)$, $\mathcal{G}_c(1;2) = 
\mathcal{G}(1;2) - \mathcal{G}(1) \mathcal{G}(2)$, etc. 
We point out that in the general case, these connected Green's functions 
have to be calculated in the presence of the $V$ and $U$ interaction terms.

Expressing $E$ in terms of $A$ and $\theta$ it is clear that no linear
terms in $A$ or $\theta$ can appear in $\mathcal{S}_{e\!f\!f}$ 
for a non-vanishing momentum $\bm{q}$. The quadratic terms in $A$ and 
$\theta$ of $\mathcal{S}_{e\!f\!f}$ become
\begin{eqnarray}
\mathcal{S}_{e\!f\!f}[A,\theta] &=& \frac{1}{2} \sum_{q,\alpha,\alpha'}
A_\alpha(q) \mathcal{K}^{AA}_{\alpha\alpha'}(q)
A_{\alpha'}(-q)\nonumber \\
& &+~ \frac{1}{2} \sum_{q} \theta(q) \mathcal{K}^{\theta\theta}(q)
\theta(-q)\nonumber \\
& &+ \sum_{q,\alpha} A_\alpha(q) \mathcal{K}^{A\theta}_{\alpha}(q)
\theta(-q)~,
\label{Seff}
\end{eqnarray}
with
\begin{eqnarray}
\mathcal{K}^{AA}_{\alpha\alpha'}(q) &=& - \frac{1}{\beta \mathrm{N_s}}
\sum_{k,k'} v_\alpha(\bm{k}) v_{{\alpha}'}(\bm{k}') \nonumber \\
& &\hspace{14mm} \times~ \mathcal{G}(n_\alpha,k,q;n_{\alpha'},k',-q) \nonumber \\
& &+\frac{2}{\beta \mathrm{N_s}} \sum_{k} m_{\alpha \alpha'}^{-1}(\bm{k}) 
(1-\delta_{\alpha 0})(1-\delta_{\alpha' 0}) \nonumber \\
& &\hspace{14mm} \times~ \mathcal{G}(3,k,0) ~,
\label{KAA}  \\
\mathcal{K}^{\theta\theta}(q) &=& - \frac{1}{\beta \mathrm{N_s}}
\sum_{k,k'} w(\bm{k},\bm{q}) w(\bm{k}',-\bm{q}) \nonumber \\
& &\hspace{14mm} \times~ \mathcal{G}(2,k,q;2,k',-q) \nonumber \\
& &+\frac{2}{\beta \mathrm{N_s}} \sum_{k} z(\bm{k},\bm{q},-\bm{q}) 
\mathcal{G}(1,k,0) ~,
\label{KTheta} \\
\mathcal{K}^{A\theta}_\alpha (q) &=& - \frac{1}{\beta \mathrm{N_s}}
\sum_{k,k'} v_\alpha(\bm{k})  w(\bm{k}',-\bm{q}) \nonumber \\
& &\hspace{14mm} \times~ \mathcal{G}(n_\alpha,k,q;2,k',-q) ~,
\label{KATheta}
\end{eqnarray}
General considerations show that the Taylor expansion of 
$\mathcal{K}^{\theta\theta}$
in powers of $q$ has no constant or linear terms, that of 
$\mathcal{K}^{A\theta}$ 
no constant terms. We therefore write
\begin{eqnarray}
\mathcal{K}^{\theta\theta}(q) &=& -\sum_{\alpha,\alpha'=0}^{3}
q_\alpha \mathcal{K}^{\theta\theta}_{\alpha\alpha'}(q) q_{\alpha'}~,
\label{K1} \\
\mathcal{K}^{A\theta}_\alpha(q) &=& -\sum_{\alpha'=0}^{3}
\mathcal{K}^{A\theta}_{\alpha\alpha'}(q) q_{\alpha'}~,
\label{K2}
\end{eqnarray}
where the functions $\mathcal{K}^{\theta\theta}_{\alpha\alpha'}(q)$ and
$\mathcal{K}^{A\theta}_{\alpha\alpha'}(q)$ approach in general finite values
for $q \rightarrow 0$. Writing 
\begin{equation}
\theta_\alpha(q) \equiv q_\alpha \theta(q)~,
\end{equation}
for $\alpha=0,...,3$, we obtain for $S_{e\!f\!f}$
\begin{eqnarray}
\mathcal{S}_{e\!f\!f}[A,\theta] &=& \frac{1}{2} \sum_{q,\alpha,\alpha'} \Big( 
A_\alpha(q) \mathcal{K}^{AA}_{\alpha\alpha'}(q)A_{\alpha'}(-q) \nonumber\\
& &\hspace{14mm}+~ \theta_\alpha (q) \mathcal{K}^{\theta\theta}_{\alpha\alpha'}(q)
\theta_{\alpha'}(-q) \nonumber\\
& &\hspace{14mm}+~ 2A_\alpha(q) \mathcal{K}^{A\theta}_{\alpha\alpha'}(q)
\theta_{\alpha'}(-q) \Big)~.~~~~~
\label{Seffcont}
\end{eqnarray}
Eq.\! (\ref{Seffcont}) shows that $\mathcal{S}_{e\!f\!f}$ actually depends
not on the phase $\theta$ itself but only on its gradients. 
The three functions $\mathcal{K}^{AA}$, $\mathcal{K}^{\theta\theta}$,
and $\mathcal{K}^{A\theta}$ are not independent from each other. To see this
we apply a gauge transformation to $\mathcal{S}_{e\!f\!f}$
\cite{Ketterson1999}
\begin{eqnarray}
A_\alpha(q) &\rightarrow& A_\alpha(q) + \mathrm{i}\chi_\alpha(q)~,
 \nonumber \\
\theta_\alpha(q)  &\rightarrow& \theta_\alpha(q) 
- 2e \chi_\alpha(q)~,
\label{gauge}
\end{eqnarray}
where both equations hold for $\alpha=0,1,2,3$ and we 
have defined $\chi_\alpha(q) \equiv q_\alpha \chi(q)$.
For the following gauge transformation it is sufficient to consider  
both $\theta$ and $\chi$ to be very small. As a result we
only deal with small phase deviations from $\theta = 0$
and a possible non-uniqueness of wave functions cannot play any role.
Invariance of $\mathcal{S}_{e\!f\!f}$ against this transformation 
yields the identities
\begin{equation}
 \mathcal{K}^{AA}_{\alpha\alpha'}(q) = -4e^2  
\mathcal{K}^{\theta\theta}_{\alpha\alpha'}(q)
= -2\mathrm{i}e  \mathcal{K}^{A\theta}_{\alpha\alpha'}(q)~.
\label{equal}
\end{equation}
Explicit calculations for the case of non-interacting electrons
are presented in the appendix and provide direct checks of 
Eq.\! (\ref{equal}). 
Using the above relations one finally finds for $\mathcal{S}_{e\!f\!f}$
\begin{eqnarray}
\mathcal{S}_{e\!f\!f}[A,\theta] &=& \frac{1}{2} \sum_{q,\alpha,\alpha'}
\bigg[A_\alpha(q)+\frac{\mathrm{i}}{2e}\theta_\alpha(q)\bigg] 
\mathcal{K}^{AA}_{\alpha\alpha'}(q) \nonumber \\
& &\hspace{10mm} \times \bigg[A_{\alpha'}(-q)+\frac{\mathrm{i}}{2e}
\theta_{\alpha'}(-q)\bigg]~. 
\label{Seffcont1}
\end{eqnarray}

The matrix $\mathcal{K}^{AA}_{\alpha\alpha'}$ connects the charge and current 
induced by an applied external potential $A$. 
Our microscopic expression, 
Eq.\! (\ref{KAA}), just represents a generalization of the usual
expression
\cite{Schrief-book} to the case with interactions between electrons. 
$\mathcal{K}^{AA}_{00}(q)$ 
is the density-density correlation function. For a charged system
it is convenient to write it in the form,
\begin{equation}
\mathcal{K}^{AA}_{00}(q) = \frac{\tilde{\mathcal{K}}^{AA}_{00}(q)}
{1+V(\bm{q})\tilde{\mathcal{K}}^{AA}_{00}(q)}~,
\label{RPA}
\end{equation}
where $\tilde{\mathcal{K}}$ is the irreducible part of $\mathcal{K}$, i.e., 
it contains all 
diagrams
to $\mathcal{K}$ which cannot be decomposed into two parts by cutting one 
Coulomb line. \cite{DiagrNambu}
Taking into account the layered structure 
of high-T$_c$ cuprates, the Coulomb potential $V(\bm{q})$
 is given by\cite{Fetter}
\begin{equation}
V(\bm{q}) = \frac{2\pi e^2d}{\epsilon_b q_{\parallel}}
\Big[ \frac{\sinh(q_\parallel d)}{\cosh(q_\parallel d)-\cos(q_z d)}
\Big]~,
\label{Coulomb}
\end{equation}
with $q_\parallel = \sqrt{q_x^2+q_y^2}$. $d$ is the distance between
layers and $\epsilon_b$ a background dielectric constant.
$\mathcal{K}^{AA}_{00}$ 
approaches at small wave vectors the universal function $1/V(\bm{q})$.
Symmetry requires $\mathcal{K}^{AA}_{0\alpha'} =0$ for $\alpha'=1,2,3$. 
Furthermore,
$\mathcal{K}^{AA}_{\alpha\alpha'}$ is for $\alpha,\alpha'=1,2,3$ already 
irreducible
in the above sense so that no analogue to Eq.\! (\ref{RPA}) exists in this
case. 

Finally we need an equation to determine $\Delta_0$. 
Using the Nambu formulation the terms in the small square brackett in 
Eq. (\ref{StJ-def}) can be written after a Fourier transformation as
\begin{eqnarray}
& &-\frac{1}{4}\int_0^\beta d \tau \sum_{\langle ij \rangle} [\Delta_{ij}
\bar{B}^*_{ij}(\tau) + \Delta_{ij}^*(\tau) \bar{B}_{ij}(\tau)] 
\nonumber \\
& &~=~ \sum_{k,q} \bar{\Phi}_k \tilde{\Delta}_{k,k+q} \Phi_{k+q}~.
\label{A1}
\end{eqnarray}
$\tilde{\Delta}_{k,k+q}$ is, except for a factor $-1/4$,
the Fourier transform of
$\Delta_{ij}(\tau)\bm{\sigma^+} + \Delta_{ij}^*(\tau) \bm{\sigma}^{-}$
with $\bm{\sigma}^{\pm} = (\bm{\sigma}_{1} \pm \bm{\sigma}_{2})/2$.
In determining $\Delta_0$ we may put all phases to zero so that
$\Delta_{ij}(\tau)\bm{\sigma}^{+} + \Delta_{ij}^*(\tau) \bm{\sigma}^{-}$
reduces to $\Delta_0 \gamma_{ij}\bm{\sigma}_{1}$. Considering the deviation
$\delta \Delta_0$ from the saddle point value $\Delta_0$ the 
corresponding change in the action on the right-hand side of Eq. (\ref{A1})
yields a change linear in $\delta \Delta_0$ in $\mathcal{S}_{eff}$ after 
integrating out the Fermi fields as previously. Requiring that
the total linear term in $\delta \Delta_0$ vanishes yields the 
desired equation 
\begin{equation}
\frac{\Delta_0}{J} = \frac{1}{\beta \mathrm{N_s}} \sum_k 
(\cos{k_x} -\cos{k_y}) \;\mathcal{G}(1,k,0).
\label{saddle}
\end{equation}

\section{Gradient expansion of $\mathcal{S}_{eff}$}

\begin{figure*}[t]
\includegraphics[width=119mm]{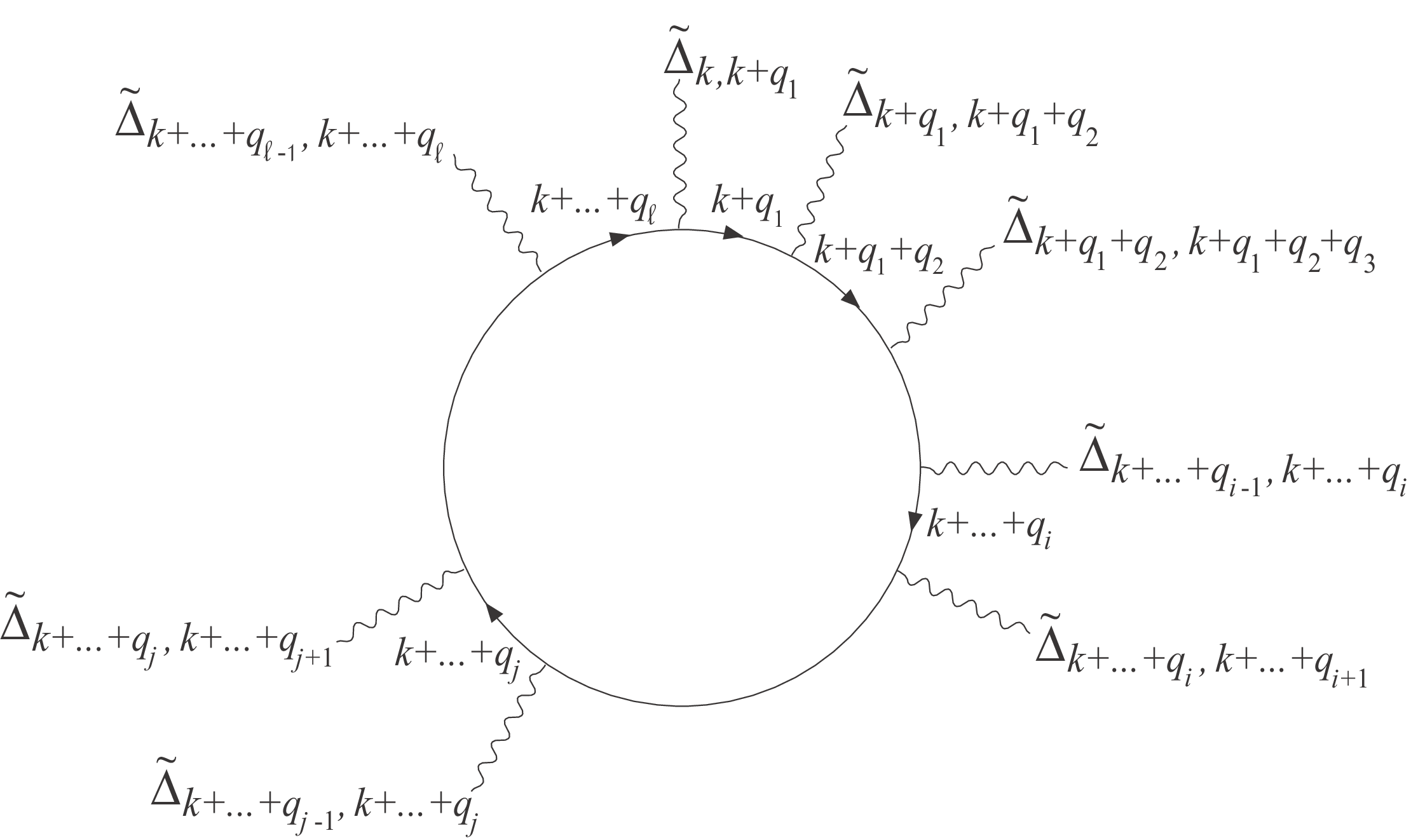}
\caption {Ring diagram of order $\ell$ in the order parameter 
$\tilde{\Delta}_{k,k+q}$. The solid lines denote the unperturbed Green's 
function ${\mathcal{G}}_{0}$.}
\label{Fig1}
\end{figure*}

$\mathcal{S}_{eff}$ has been derived in section III under the
assumption that the phase $\theta$ is small. An expansion
of $\mathcal{S}_{eff}$ in terms of gradients of the order parameter
seems to be more satisfying because $\theta$ is then no longer
restricted to small values.
As shown in Ref. \onlinecite{Benedetti} for the case of a charge-density-wave
state, resummations allow to transform the expansion of $\mathcal{S}_{eff}$
in powers of the order parameter into an expansion in powers of gradients
of the order parameter. In the following we adapt this method to our action
and consider first the non-interacting case $\mathcal{S}_{int}=0$.
The integration over fermions can then 
easily be performed and one obtains, dropping the constant 
$\sim \Delta_0^2$,
\begin{eqnarray}
\label{A2}
\mathcal{S}_{e\!f\!f}[0,\theta] &=& -\mathrm{Tr} \Big\{ 
\mathrm{ln}(-{\mathcal{G}}_{0}^{-1}+\tilde{\Delta})
\Big\}  \\
&=& -\mathrm{Tr}\Big\{ \mathrm{ln} 
\big(-{\mathcal{G}}_{0}^{-1}\big)\Big\} +\sum_{\ell=1}^\infty
\frac{1}{\ell} \mathrm{Tr}\Big\{ ({\mathcal{G}}_{0} \tilde{\Delta})^{\ell} 
\Big\}~. \nonumber 
\end{eqnarray}
$\mathcal{G}_0$ is the unperturbed Green's function due to the two first
terms in Eq.\! (\ref{Sdwave-def}), its expression is given in 
Eq.\! (\ref{Green0-def}). $\mathrm{Tr}$ denotes the trace over $k$ and the
Nambu index. The $\ell$-th order term $X_{\ell}$ in the sum over $\ell$ 
can be written in frequency-momentum space as
\begin{eqnarray}
& & X_\ell = \frac{1}{\ell} \sum_{q_1,...,q_\ell} \int dr~
e^{\mathrm{i}r\cdot(q_1+...+q_\ell)} Y(q_1,q_2,...,q_\ell)~,~~~~~~
\label{Xl} \\
& & Y(q_1,q_2,...,q_\ell) = \sum_k
\mathrm{Tr'} \Big\{ \tilde{\Delta}_{k,k+q_1} 
{\mathcal{G}}_{0}(k+q_1) ~ ...
\nonumber \\
& &\hspace{38mm} \times~ \tilde{\Delta}_{k+...q_{\ell-1},k+...+q_\ell} 
\nonumber \\
& &\hspace{38mm} \times~ {\mathcal{G}}_{0}(k+...+q_\ell) \Big\}~.
\label{A3}
\end{eqnarray}
$r$ stands for the vector $(\tau,{\bm{r}}_i)$, and $\int dr$ for
$1/(\beta \mathrm{N_s})\sum_i \int_0^\beta d\tau$. $\mathrm{Tr'}$ 
denotes a trace over the Nambu index.
$X_\ell$ can be visualized by a ring diagram (see Fig.\! \ref{Fig1}) where
the electronic Green's function ${\mathcal{G}}_{0}$ (solid line)
is scattered at
the external potentials $\tilde{\Delta}$ (wavy lines). 

For the following it is
convenient to write $\tilde{\Delta}(k,q_1)$ instead of 
$\tilde{\Delta}_{k,k+q_1}$. The first momentum $k$ refers then to
the relative and the second one $q_1$ to the center-of-mass
motion of the Cooper pair. $k$ may assume arbitrary values whereas
$q_1,q_2,...,$ are considered to be small. 
\begin{figure}[b]
\includegraphics[width=81mm]{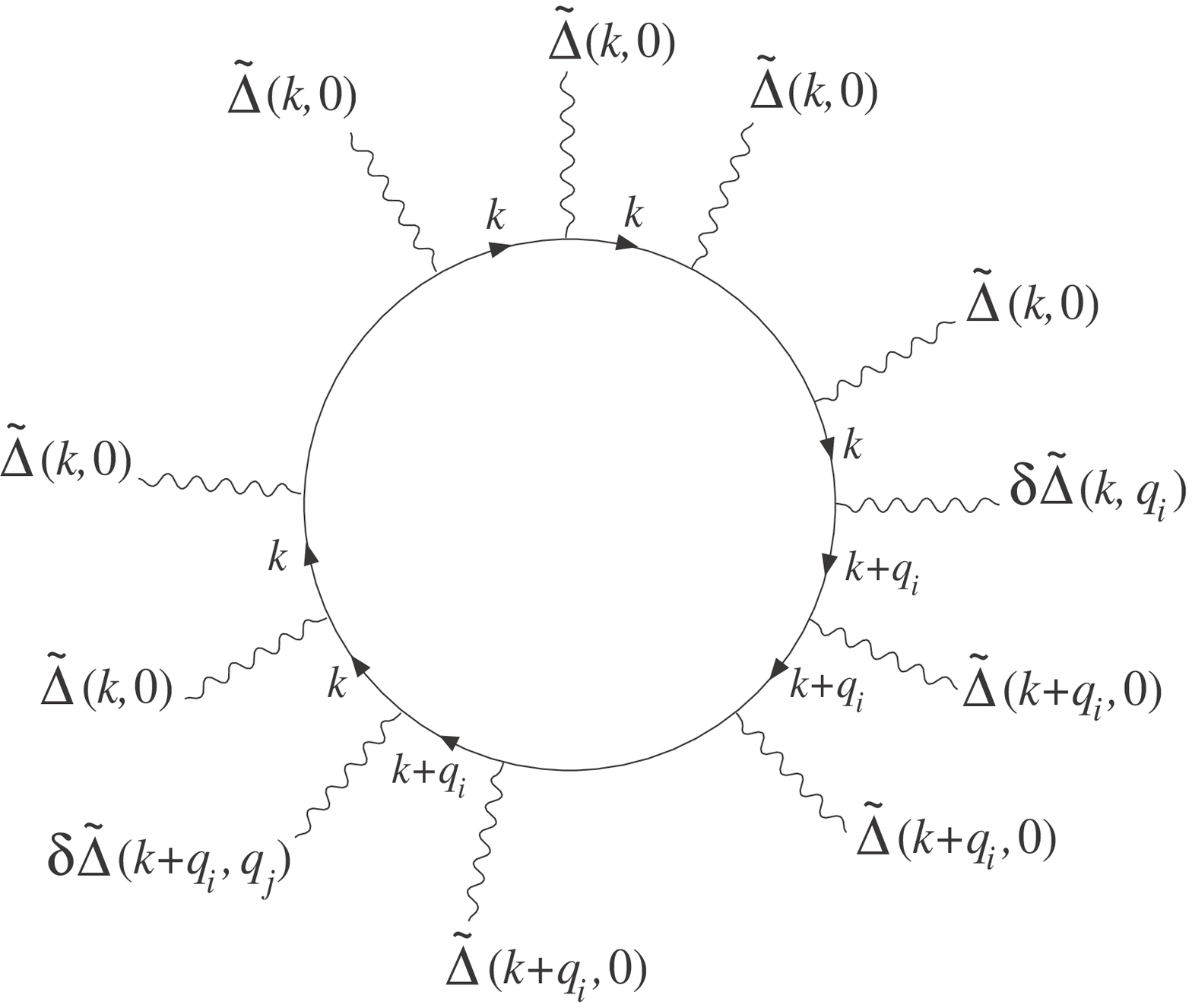}
\caption {Ring diagram of order $\ell$ in the order parameter 
$\tilde{\Delta}_{k,k+q}$ with momenta corresponding to small phase
fluctuations around an equilibirum state with long-range order.
}
\label{Fig2}
\end{figure}

\begin{figure}[t]
\includegraphics[width=85mm]{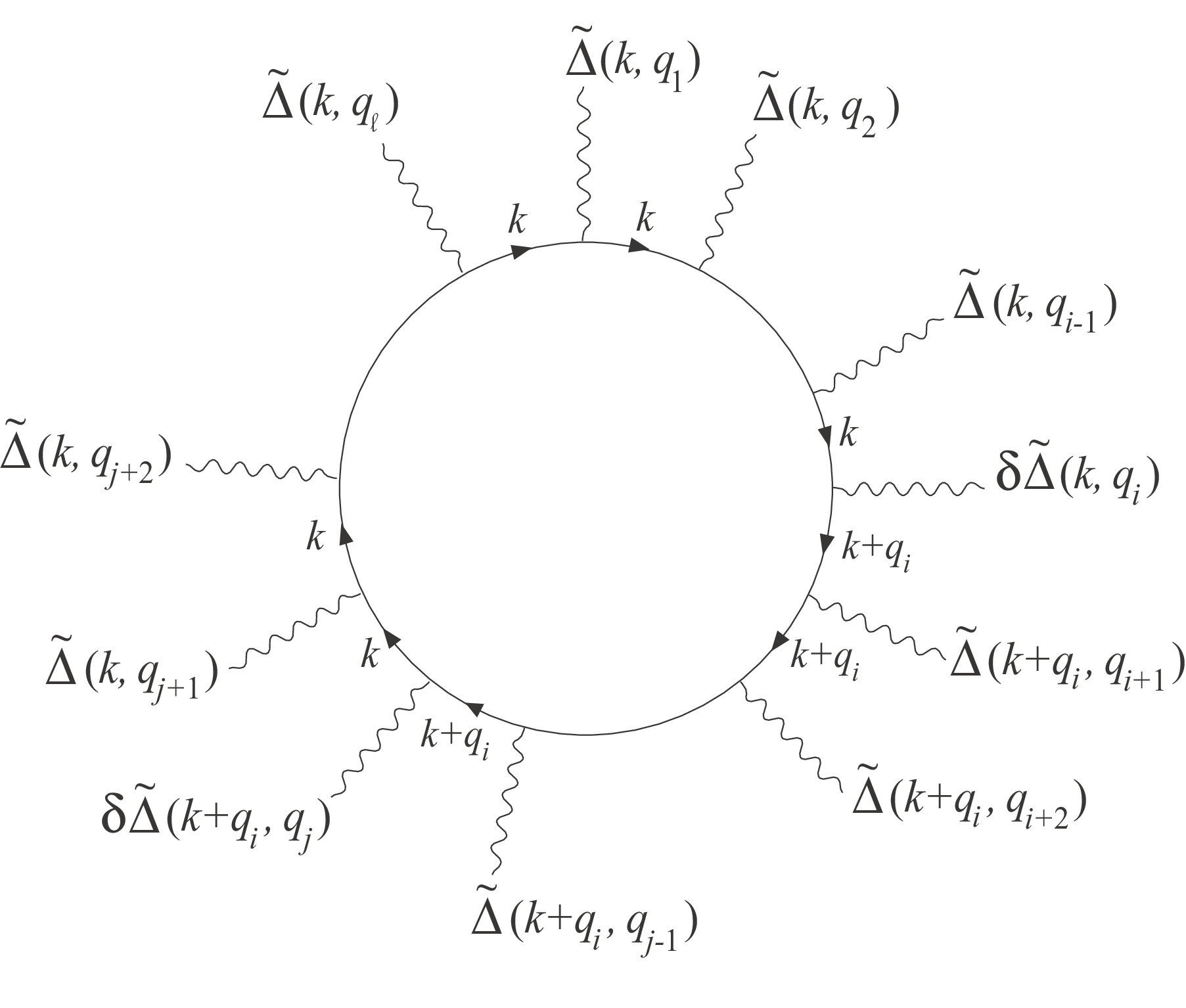}
\caption {Ring diagram of order $\ell$ in the order parameter 
$\tilde{\Delta}_{k,k+q}$ with momenta corresponding to the leading
non-local contribution in the gradient expansion.
}
\label{Fig3}
\end{figure}

In the non-interacting case the calculations for $\mathcal{S}_{eff}$
in section III correspond to the evaluation of ring diagrams of the type shown
in Fig.\! \ref{Fig1}. To obtain $\mathcal{S}_{eff}$ in second order in $\theta$
we had to take into account one and also two non-equilibrium
external lines due to phase fluctuations yielding 
the second and first terms in Eq.\! (\ref{KTheta}), respectively. 
The corresponding ring diagram of order $\ell$ is obtained by
evaluating the Green's functions in Eqs.\! (\ref{KAA})-(\ref{KATheta})
for non-interacting electrons and expanding them to order $\ell -2$
or $\ell -1$, respectively, in $\tilde{\Delta}(k,0)$.
Let us first consider 
the first term in more detail. Its $\ell$-th order contribution
can be illustrated by the diagram shown in Fig.\! \ref{Fig2}. 
$\delta \tilde{\Delta}(k,q_i)$
and $\delta \tilde{\Delta}(k+q_i,q_j)$ are the two non-equilibrium
lines with $q_i=-q_j$ from momentum conservation. Between these lines
the electrons are $\ell -2$ times scattered at the equilibrium
order parameter 
with zero momentum. In this case the phase deviations from their
equilibrium values must be considered as small so that the expansion
in powers of $\theta$ is appropriate.
If the phase does not exhibit true long-range order but still varies
slowly in space and time the ring diagrams of Fig.\! \ref{Fig1}  have
to be evaluated in a different way. In this case 
the momenta $q_i$ in the external lines have to be kept 
but one may expand the electron 
propagators in powers of $q_i$. As shown in 
Refs.\! \onlinecite{Benedetti} and \onlinecite{Muramatsu} such an expansion
generates a gradient expansion for $\mathcal{S}_{eff}$. 
In particular, it does not assume that $\theta$ is small
but only that the gradients of $\theta$ are small. Fig.\! \ref{Fig3} shows the 
distribution of momenta which yield the leading term in this 
expansion. Scattering at the sites
$i$ and $j$ is accompagnied by a change in energy and momentum $q_i$,
whereas no change in energy or momentum occurs at all the other sites.
Keeping higher terms in the expansion of the electron
propagator in terms of $q_i$ 
would yield contributions to $\mathcal{S}_{eff}$ which are
at least of third order in phase gradients.     

Evaluating the diagram of Fig.\! \ref{Fig3}
the sums over $q_1, ..., q_{i-1}, q_{i+1}, ..., q_{j-1}, q_{j+1} , ... q_\ell$ can be
immediately be carried out yielding products of $\tilde{\Delta}(k,r)$, 
the Fourier transform of $\tilde{\Delta}(k,k_1)$ with respect to $k_1$.   
It is convenient to introduce a Green's function 
$\tilde{\mathcal{G}}(k,r)$ by
\begin{equation}
\tilde{\mathcal{G}}^{-1}(k,r) = \mathcal{G}_0^{-1}(k) -
\tilde{\Delta}(k,r)~.
\label{A4}
\end{equation} 
After a small rearrangement of terms one obtains for the diagram
the expression
\begin{eqnarray}
X_{\ell}^{(2)} &=& \frac{1}{2\ell} \sum_{i\neq j} \sum_{k,q_i} \int dr~
\mathrm{Tr'}\Big\{ \tilde{\mathcal{G}}^{(\ell-j+i-1)}(k,r) \tilde{\Delta}(k,q_i)
\nonumber \\
& &\hspace{30mm} \times~ \tilde{\mathcal{G}}^{(j-i-1)}(k+q_i,r)
 \nonumber \\
& &\hspace{30mm} \times~
 \tilde{\Delta}(k,-q_i)\Big\}~,
\label{A6}
\end{eqnarray}
where $\tilde{\mathcal{G}}^{(\ell)}(k,r)$ is the contribution of order $\ell$
to $\tilde{\mathcal{G}}$. The sums over $i,j$ and finally also over $\ell$
can also be performed yielding the following contribution 
to $\mathcal{S}_{eff}$
\begin{eqnarray}
\mathcal{S}_{eff}^{(2)} &=& \frac{1}{2} \sum_{k,q} \int dr~
\mathrm{Tr'}\Big\{ \tilde{\mathcal{G}}(k,r) \tilde{\Delta}(k,q)
 \nonumber \\
& &\hspace{24mm} \times~ \tilde{\mathcal{G}}(k+q,r)\tilde{\Delta}(k,-q) \Big\}~.~~~~~
\label{A7}
\end{eqnarray}

Ring diagrams with only one non-equilibrium line 
$\delta \tilde{\Delta}(k,q)$ can be evaluated in a similar manner. 
Though these diagrams only involve the $q=0$ Fourier component
of $\tilde{\Delta}$ due to energy and momentum conservation
they yield products of phase fluctuations when passing from 
order parameter to phase fluctuations. 
Evaluating these ring diagrams in form of a gradient expansion yields the 
following contribution to $\mathcal{S}_{eff}$
\begin{equation}
\mathcal{S}_{eff}^{(1)} = \sum_k \int dr~
\mathrm{Tr'}\Big\{ \tilde{\mathcal{G}}(k,r) 
\tilde{\Delta}(k,0)\Big\}~.
\label{A8}
\end{equation}

Eqs.\! (\ref{A7}) and (\ref{A8}) represent time and space averages of the 
corresponding homogenous action where the Green's functions contain the 
local instead of the global gap. This gives a simple recipe to generalize
an expression for $\mathcal{S}_{eff}$ derived under the assumption of small 
phase fluctuations to one which is valid also for large phase fluctuations
in the leading order of a gradient expansion: one writes $\mathcal{S}_{eff}$
as a density in space and time and then uses at a given point $r$
in space and time the homogenous expression for $\mathcal{S}_{eff}$
with the constant gap value $\tilde{\Delta}(k,r)$ in the Green's functions.
\indent In the phase-only approximation order parameter fluctuations
are solely due to fluctuations in the phase. Because the Green's functions
in Eq.\! (\ref{A7}) refer for a given $r$ to a constant phase 
this phase can be gauged away by a global gauge transformation
without any additional contribution to the vector potential $A$.
The right-hand side of Eq.\! (\ref{A7}) is clearly invariant against 
a global gauge transformation. This is true even separatly for the product
of the order parameters and the susceptibility because the latter involves
the same number of creation and annihilation operators so that global
phases cancel. In the phase-only
approximation we thus may put the phase in Eq.\! (\ref{A7}) to zero.
This means that the ring diagram of Fig.\! \ref{Fig4} reduces to a
diagram of second-order in $\tilde{\Delta}$ where the Green's functions
(solid lines) are to be calculated with the equilibrium order parameter
Eqs.\! (\ref{Delta-gen}) and (\ref{Delta-dwaveEq}). This result proves 
that within the phase-only approximation       
the effective action derived for small fluctuations in section III
is in leading order in field gradients identical with that of the gradient 
expansion.
In the general case where also the amplitude $\Delta_0$ varies in time
and space Eq.\! (\ref{A7}) clearly differs from the corresponding lowest-order 
expression in section III.
   
It is evident that the above results also hold for 
interacting electrons. For a given skeleton diagram in interaction
and electron lines there is again a one-to-one correspondence between 
diagrams in $\ell$-th order in $\tilde{\Delta}$ for 
ordered and disordered ground states. For ordered ground states
this diagram is obtained by expanding electron lines in 
$\tilde{\Delta}(k,0)$ so that the total order is $\ell -2$. 
In the disordered case the same diagram is obtained by expanding
$\mathcal{S}_{eff}$ in Eq.\! (\ref{Seff-NegOr}) up to the order $\ell$,
and associating in all possible ways two vertices with 
and the remaining $\ell -2$ without energy and momentum changes.
Writing the energy and momentum conservation as a Fourier integral
over $r$, the momentum integration at the $\ell-2$ vertices without
energy and momentum changes can be carried out yielding products
of $\tilde{\Delta}(k,r)$. As a result a strict one-to-one correspondence
between diagrams of the ordered and disordered cases is established 
from which the above recipe follows.

\section{Discussion and Conclusions}

\begin{figure}[t]
\includegraphics[width=81mm]{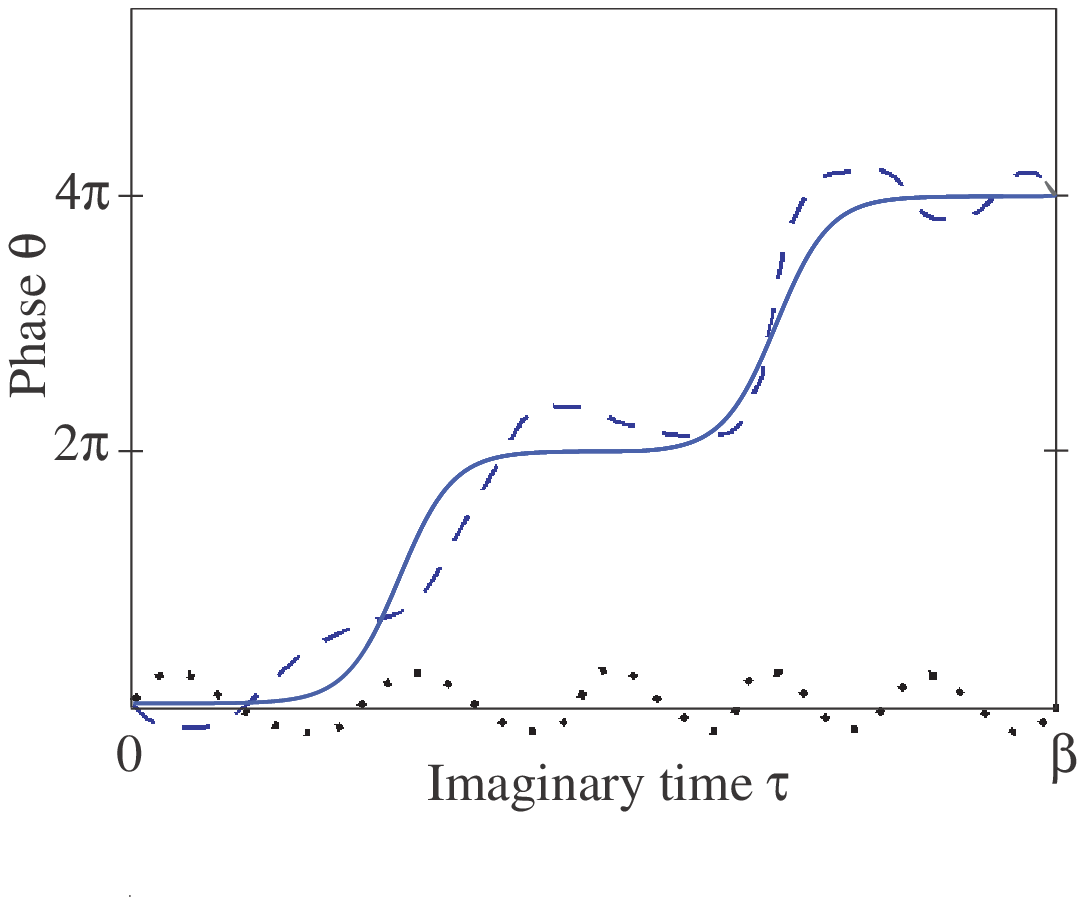}
\caption {(Color online) Small and large fluctuations in the $\theta$-$\tau$ plane.}
\label{Fig4}
\end{figure}

Eq.\! (\ref{Seffcont1}) together with 
Eqs.\! (\ref{KAA})-(\ref{K2}) represent a microscopic 
expression for the effective action of phase fluctuations
in a $d$-wave superconductor with interacting electrons.
In section III this expression was derived under the assumption
that $\theta$ is small. The dotted line in Fig.\! \ref{Fig4}
illustrates this case for the $\tau$-dependence of $\theta$.
Within the interval $[0,\beta]$ $\theta$ performs
small oscillations around the equilibrium value $\theta = 0$. 
Similar pictures can be drawn for paths in $r=(\tau,\bm{r})$
space.  
The gradient expansion in section IV allows to consider  more general 
paths of the form $\theta(r)= \theta_0(r) + \delta \theta(r)$. 
$\theta_0(r)$ is assumed to be slowly varying with $r$ whereas 
$\delta \theta(r)$ must be small, as illustrated by the solid
and dashed lines, respectively, in Fig.\! \ref{Fig4}. Since
in the phase-only approximation the gradient expansion leads to
the same expression 
for $\mathcal{S}_{eff}$ as the perturbation theory
Eq.\! (\ref{Seffcont1}) is actually valid for all paths shown
in Fig. \ref{Fig4}. To determine
$\mathcal{S}_{eff}$ for the solid curve 
it is sufficient to take in Eq.\! (\ref{Seffcont1}) the long-wavelength,
low-frequency limit in $K^{AA}_{\alpha\alpha'}(q)$.
The important slowly varying extremal paths satisfy then a  
second-order differential equation which may have 
besides of trivial constant also vertex solutions 
illustrated by the solid line in Fig.\! \ref{Fig4}. 

$\mathcal{S}_{eff}$ reduces in the static limit to the
Ginzburg-Landau form for the phase-dependent part of the free energy of a 
superconductor. For non-interacting electrons and $A=0$ it also coincides
with the expression given in Ref.\! \onlinecite{Param-PRB2000}. 
In contrast to the latter reference 
the expression of $\mathcal{S}_{eff}$ we have obtained,
displayed in Eq.\! (\ref{Seffcont1}), is explicitly gauge-invariant
due to the inclusion of $A$. 
Another reason why $A$ should be kept in $\mathcal{S}_{eff}$ is
that current fluctuations produce a magnetic field which cannot be gauged away.
Heuristically, the above  expression for $\mathcal{S}_{eff}$ can be 
obtained in the following simple way: Discard order parameter fluctuations 
in the original 
microscopic Lagrangian but keep the potential $A$. Integrate then 
over the Fermi fields and make the resulting $\mathcal{S}_{eff}$ 
gauge-invariant by applying the gauge 
transformation, Eq.\! (\ref{gauge}), and interprete the  
field $-2e\chi$ as the phase field of the order parameter. Finally, we stress
that our derivation of $\mathcal{S}_{eff}$ did not make use of any singular
gauge transformation as in Refs.\! \onlinecite{Param-PRB2000} and \onlinecite{Carbotte2004}.
According to Refs.\! \onlinecite{FrTesa-PRL2001,FrTesa-PRB2002,VaTesa-PRL2003} 
the use of such a transformation
is problematic because it produces in general additional cut contributions. 
The fact that our $\mathcal{S}_{eff}$ agrees
for $A=0$ and non-interacting electrons with that of Ref.\! \onlinecite{Param-PRB2000},
however, suggests that such cut contributions must cancel out in 
$\mathcal{S}_{eff}$ and that a simple use of singular gauge transformations 
leads to the correct result for $\mathcal{S}_{eff}$. 

\begin{acknowledgements}

We are grateful to Dirk Manske for a careful reading of the manuscript.

\end{acknowledgements}

\appendix

\section{Interrelations between different kernels $\mathcal{K}$ in 
$\mathcal{S}_{e\!f\!f}$ for non-interacting electrons}
\label{Appen-FreeEl}

In this appendix we check the relation between the kernels 
$\mathcal{K}^{AA}$ and $\mathcal{K}^{\theta\theta}$ 
given by Eq.\! (\ref{equal}) for an isolated layer and non-interacting
electrons. The free Green's function matrix is then given by
\begin{eqnarray}
{\mathcal G}_{0}(k) &=&
\frac{-1}{(\nu_{m})^{2}+(\xi_{\bm{k}})^{2}+(\Delta_{\bm{k}})^{2}}
\nonumber \\
& & \times~
\left[ \begin{array}{cc}
\mathrm{i} \nu_{m} + \xi_{\bm{k}} \ & \ -\Delta_{\bm{k}} \\
-\Delta_{\bm{k}} \ & \ \mathrm{i} \nu_{m} - \xi_{\bm{k}}
\end{array} \right]~,
\label{Green0-def}
\end{eqnarray}
with the following energies
\begin{eqnarray}
 \label{xiEnQuant-def}
\Delta_{\bm{k}} &=& \frac{\displaystyle \Delta_{0}}{\displaystyle 2} 
\big[\cos({k}_{x})-\cos({k}_{y})\big]~,~ \xi_{\bm{k}} = \epsilon_{\bm{k}} - \mu~,
 \\
\epsilon_{\bm{k}} &=& 
-2t \big[\cos({k}_{x})+\cos({k}_{y})\big] - 
4t^{'}\big[\cos({k}_{x}).\cos({k}_{y})\big]~,
 \nonumber 
\label{DelEnQuant-def}
\end{eqnarray}
where $\epsilon_{\bm{k}}$ includes nearest and next-nearest 
neighbor hopping contributions.
Eq. (\ref{saddle}) yields the BCS gap equation~\cite{BCS,Schrief-book}
\begin{eqnarray}
\frac{1}{J} = \frac{1}{\mathrm{N_s}} \sum_{\bm{k}} 
\frac{\big[\cos({k}_{x})-\cos({k}_{y})\big]^{2}}{2 E_{\bm{k}}}
\mathrm{tanh} \bigg(\frac{\beta E_{\bm{k}}}{2} \bigg)~,~~
\label{SadP-eq}
\end{eqnarray}
\vfill\eject

\noindent where $E_{\bm{k}}$ is the quasiparticle energy, $E_{\bm{k}} = 
\sqrt{(\xi_{\bm{k}})^{2}+(\Delta_{\bm{k}})^{2}}$, and the summation in 
$\bm{k}$-space is extended over the first (square) Brillouin zone.
Performing the summation over the fermionic Matsubara frequencies
we find for the kernel 
$\mathcal{K}^{\theta\theta}$ of Eq. (\ref{KTheta}),

\begin{widetext}
\begin{eqnarray}
& &\mathcal{K}^{\theta\theta}(q)
= \frac{1}{4\mathrm{N_{s}}} \sum_{\bm{k}} \Bigg(
\big\{ \Delta_{\bm{k}}+\Delta_{\bm{k}-\bm{q}} \big\}^{2}
\Bigg[ \frac{\big\{1-n_{\mathrm{FD}}(E_{\bm{k}-\bm{q}})
-n_{\mathrm{FD}}(E_{\bm{k}}) \big\}}{2}
\bigg\{1 + \frac{\xi_{\bm{k}} \xi_{\bm{k}-\bm{q}}+\Delta_{\bm{k}} 
\Delta_{\bm{k}-\bm{q}} }{E_{\bm{k}} E_{\bm{k}-\bm{q}} } \bigg\} 
 \nonumber \\
& &\hspace{103mm} \times~ \bigg\{ \frac{1}{\textrm{i}\omega_{n} - E_{\bm{k}} 
- E_{\bm{k}-\bm{q}}}
- \frac{1}{\textrm{i}\omega_{n} + E_{\bm{k}} + E_{\bm{k}-\bm{q}}} \bigg\} ~~
\nonumber \\
& &\hspace{55mm} +~\frac{\big\{n_{\mathrm{FD}}(E_{\bm{k}-\bm{q}})
-n_{\mathrm{FD}}(E_{\bm{k}}) \big\}}{2}
\bigg\{1 - \frac{\xi_{\bm{k}} \xi_{\bm{k}-\bm{q}}+\Delta_{\bm{k}} 
\Delta_{\bm{k}-\bm{q}} }{E_{\bm{k}} E_{\bm{k}-\bm{q}} } \bigg\} 
\nonumber \\
& &\hspace{99mm} \times~ \bigg\{ \frac{1}{\textrm{i}\omega_{n} - E_{\bm{k}} 
+ E_{\bm{k}-\bm{q}}}
- \frac{1}{\textrm{i}\omega_{n} + E_{\bm{k}} - E_{\bm{k}-\bm{q}}} \bigg\} 
\Bigg]
\nonumber \\
& & \hspace{32mm} +~ \frac{2 \Delta_{\bm{k}} \big\{ 
\Delta_{\bm{k}}+\Delta_{\bm{k}-\bm{q}} \big\}}{E_{\bm{k}}}
\big\{1 - 2 n_{\mathrm{FD}}(E_{\bm{k}}) \big\} \Bigg)~,
\label{Kern2D-full}
\end{eqnarray}
\end{widetext}
with $n_{\mathrm{FD}}$ the Fermi-Dirac distribution function.

To perform a first check of  Eq.\! (\ref{equal}), we consider the frequency 
dependence of the kernels $\mathcal{K}^{\theta\theta}$ and $\mathcal{K}^{AA}$. 
With Eqs.\! (\ref{K1}) and (\ref{Kern2D-full}) we have
\begin{eqnarray}
\mathcal{K}^{\theta\theta}_{00}(\textrm{i}\omega_{n}, \bm{0})
&=& -\frac{1}{\mathrm{N_{s}}} \sum_{\bm{k}} \big\{1 - 2 
n_{\mathrm{FD}}(E_{\bm{k}}) \big\}
\nonumber \\
& & \hspace{12mm} \times~
\frac{(\Delta_{\bm{k}})^{2}}{E_{\bm{k}} \big\{ (\textrm{i}\omega_{n})^{2} 
- 4 (E_{\bm{k}})^{2} \big\}}~.~~~~~
\label{KTheta_00}
\end{eqnarray}
Using Eq.\! (\ref{KAA}) we get for the electromagnetic field kernel
\begin{eqnarray}
\mathcal{K}^{AA}_{00}(\textrm{i}\omega_{n}, \bm{0}) = 
\frac{e^2}{\beta \mathrm{N_s}} 
\sum_{k} \mathrm{Tr'}\Big[ {\bm{\sigma}}_{3} {\mathcal G}_{0}(k) 
{\bm{\sigma}}_{3} {\mathcal G}_{0}(k) \Big] ~,
\label{KAA_00-Ini}
\end{eqnarray}
which gives after summation over the fermionic frequencies
\begin{eqnarray}
\mathcal{K}^{AA}_{00}(\textrm{i}\omega_{n}, \bm{0}) &=&
\frac{4 e^2}{\mathrm{N_s}} \sum_{\bm{k}} \big\{1 - 
2 n_{\mathrm{FD}}(E_{\bm{k}}) \big\}
\nonumber \\
& & \hspace{11mm} \times~
\frac{(\Delta_{\bm{k}})^{2}}{E_{\bm{k}} \big\{ (\textrm{i}\omega_{n})^{2} 
- 4 (E_{\bm{k}})^{2} \big\}}~.~~~~~
\label{KAA_00-Fin}
\end{eqnarray}
By comparing Eqs\! (\ref{KTheta_00}) and (\ref{KAA_00-Fin}) one can see 
immediatly
\begin{eqnarray}
\mathcal{K}^{AA}_{00}(\textrm{i}\omega_{n}, \bm{0}) = -4e^2  
\mathcal{K}^{\theta\theta}_{00}(\textrm{i}\omega_{n}, \bm{0})~,
\label{Equal-Freq}
\end{eqnarray}
in agreement with Eq.\! (\ref{equal}), for any frequency 
$\textrm{i}\omega_{n}$.

The low-frequency, long-wavelength limit of the effective action 
${\mathcal{S}}_{e\!f\!f}[0,\theta]$ can be investigated by expanding 
quadratically its kernel. We start by considering the zero-temperature case. 
The low-frequency expansion up to the second-order in $\textrm{i}\omega_{n}$ 
of $\mathcal{K}^{\theta\theta}$, Eq. (\ref{Kern2D-full}), is straightforward, 
we have
\begin{eqnarray}
\mathcal{K}^{\theta\theta}(\textrm{i}\omega_{n} \rightarrow 0,\bm{q} = \bm{0})
= - \frac{1}{4 \mathrm{N_s}} \sum_{\bm{k}} \frac{(\Delta_{\bm{k}})^{2}}
{(E_{\bm{k}})^{3}} (\textrm{i}\omega_{n})^{2} ~.~~~
\label{KTTLowf}
\end{eqnarray}
However, the low-momentum expansion up to the second-order in the $\bm{q}$ 
components requires lengthy algebra. From Eq.\! (\ref{Kern2D-full}) we get

\begin{eqnarray}
\label{KTTLowq-Ini}
& & \mathcal{K}^{\theta\theta}(\textrm{i}\omega_{n} = 0,\bm{q} 
\rightarrow \bm{0})  \\
&=& \frac{1}{4 \mathrm{N_s}} \sum_{\bm{k}} \sum_{\alpha = x,y} 
(q_{\alpha})^{2} \nonumber \\
& &\hspace{1mm} \times~ \Bigg[ \frac{\xi_{\bm{k}}(\Delta_{\bm{k}})^{2}}
{(E_{\bm{k}})^{3}} 
\bigg(\frac{{\partial}^{2} \xi_{\bm{k}}}{\partial k_{\alpha}^{2}}\bigg)
- \frac{(\xi_{\bm{k}})^{2}\Delta_{\bm{k}}}{(E_{\bm{k}})^{3}}
\bigg(\frac{{\partial}^{2} \Delta_{\bm{k}}}{\partial k_{\alpha}^{2}}\bigg) 
\nonumber \\
& &\hspace{5mm} +~ \frac{2 (\Delta_{\bm{k}})^{2} - 
(\xi_{\bm{k}})^{2}}{(E_{\bm{k}})^{5}}
\bigg\{ (\Delta_{\bm{k}})^{2} \bigg(\frac{\partial \xi_{\bm{k}}}
{\partial k_{\alpha}}\bigg)^{2} 
\nonumber \\
& &\hspace{33mm} +~ (\xi_{\bm{k}})^{2} \bigg(\frac{\partial \Delta_{\bm{k}}}
{\partial k_{\alpha}}\bigg)^{2} 
\nonumber \\
& &\hspace{33mm} -~ 2 \xi_{\bm{k}} \Delta_{\bm{k}} \bigg(\frac{\partial 
\xi_{\bm{k}}}{\partial k_{\alpha}}\bigg)
\bigg(\frac{\partial \Delta_{\bm{k}}}{\partial k_{\alpha}}\bigg) \bigg\} 
\Bigg]~. \nonumber
\end{eqnarray}
The previous expression can be simplified by making use of the BCS equation, 
Eq.\! (\ref{SadP-eq}), and its second derivatives with respect to 
$k_{\alpha}$. This yields
\begin{eqnarray}
\label{KTTLowq-Interm}
& & \mathcal{K}^{\theta\theta}(\textrm{i}\omega_{n} = 0,\bm{q} 
\rightarrow \bm{0})  \\
&=& \frac{1}{4 \mathrm{N_s}} \sum_{\bm{k}} \sum_{\alpha = x,y} 
(q_{\alpha})^{2} \nonumber \\
& &\hspace{1mm} \times \Bigg[ \frac{\Delta_{\bm{k}}}{E_{\bm{k}}}
\bigg(\frac{{\partial}^{2} \Delta_{\bm{k}}}{\partial k_{\alpha}^{2}}\bigg) 
+  \frac{(\Delta_{\bm{k}})^{2}}{(E_{\bm{k}})^{3}} \bigg(\frac{\partial 
\xi_{\bm{k}}}{\partial k_{\alpha}}\bigg)^{2} 
\nonumber \\
& &\hspace{5mm} +~ \frac{(\xi_{\bm{k}})^{2}}{(E_{\bm{k}})^{3}} \bigg(\frac{\partial 
\Delta_{\bm{k}}}{\partial k_{\alpha}}\bigg)^{2} 
-2 \frac{\xi_{\bm{k}} \Delta_{\bm{k}}}{(E_{\bm{k}})^{3}} 
\bigg(\frac{\partial \xi_{\bm{k}}}{\partial k_{\alpha}}\bigg)
\bigg(\frac{\partial \Delta_{\bm{k}}}{\partial k_{\alpha}}\bigg) 
\Bigg] ~. \nonumber
\end{eqnarray}
We know that in the case of an infinite two-dimensional square lattice, 
the summation over $\bm{k}$ in the first Brillouin zone can be computed by 
performing the replacement
\begin{eqnarray}
& &\frac{1}{\mathrm{N_s}} \sum_{\bm{k}} 
\longrightarrow \frac{1}{(2 {\pi})^{2}} \int_{-\pi }^{+\pi} d k_{x} d k_{y} ~, 
\nonumber 
\end{eqnarray}
therefore we have the identity
\begin{eqnarray}
\frac{1}{\mathrm{N_s}} \sum_{\bm{k}} \bigg(\frac{{\partial}^{2} 
E_{\bm{k}}}{\partial k_{\alpha}^{2}}\bigg) = 0~.
\label{Int-0}
\end{eqnarray}
By expressing the second order derivative of $E_{\bm{k}}$ in terms of 
$\xi_{\bm{k}}$ and $\Delta_{\bm{k}}$, one can use the previous equation to 
simplify once more $\mathcal{K}^{\theta\theta}$, Eq.\! (\ref{KTTLowq-Interm}). 
We obtain
\begin{eqnarray}
\label{KTTLowq-Fin}
& &\mathcal{K}^{\theta\theta}(\textrm{i}\omega_{n} = 0,\bm{q} 
\rightarrow \bm{0})  
\\
&=& -~\frac{1}{4 \mathrm{N_s}} \sum_{\bm{k}} \sum_{\alpha = x,y} 
\frac{\xi_{\bm{k}}}{E_{\bm{k}}}
\bigg(\frac{{\partial}^{2} \xi_{\bm{k}}}{\partial k_{\alpha}^{2}}\bigg) 
(q_{\alpha})^{2} ~. 
\nonumber
\end{eqnarray}
By combining Eqs.\! (\ref{KTTLowf}) and (\ref{KTTLowq-Fin}), 
the low-frequency, long-wavelength limit of 
$\mathcal{K}^{\theta\theta}$ can finally be expressed as followed \cite{Rama-1989}
\begin{eqnarray}
\label{KTT-LowExp}
& &\mathcal{K}^{\theta\theta}(q\rightarrow 0)
\\
&=& \frac{1}{4}\Bigg\{ \chi_{0} (\textrm{i}\omega_{n} \rightarrow 0,\bm{q} 
= \bm{0}) \cdot (\textrm{i}\omega_{n})^{2}
\nonumber \\
& &\hspace{4mm} +~ \sum_{\alpha = x,y} \Lambda_{0}^{\alpha}(\textrm{i}\omega_{n} 
= 0,\bm{q} \rightarrow \bm{0}) \cdot (q_{\alpha})^{2}
\Bigg\}~.
\nonumber 
\end{eqnarray}

A similar procedure can be repeated in the finite temperature case. The 
low-frequency, long-wavelength limit of $\mathcal{K}^{\theta\theta}$ is 
given again by Eq.\! (\ref{KTT-LowExp}), with the mean field density-density 
correlation function $\chi_0$ and mean field phase stiffness 
$\Lambda_0^\alpha$, 
\begin{eqnarray}
\label{MFDeCorrFu-LOm}
& &\chi_{0} (\textrm{i}\omega_{n} \rightarrow 0,\bm{q} = \bm{0}) 
\\
&=&
-~ \frac{1}{\mathrm{N_s}} \sum_{\bm{k}} \big\{1 - 2 n_{\mathrm{FD}}
(E_{\bm{k}}) \big\}
\frac{(\Delta_{\bm{k}})^{2}}{(E_{\bm{k}})^{3}}  ~,
\nonumber
\end{eqnarray}
\begin{eqnarray}
& &\Lambda_{0}^{\alpha}(\textrm{i}\omega_{n} = 0,\bm{q} \rightarrow \bm{0})
\label{MFPhasStif-LWa} \\
&=& - ~\frac{1}{\mathrm{N_s}} \sum_{\bm{k}} 
\Bigg[ \big\{1 - 2 n_{\mathrm{FD}}(E_{\bm{k}}) \big\} 
\frac{\xi_{\bm{k}}}{E_{\bm{k}}}
\bigg(\frac{{\partial}^{2} \xi_{\bm{k}}}{\partial k_{\alpha}^{2}}\bigg)
\nonumber \\
& &\hspace{14mm} +~ 2 \beta \cdot n_{\mathrm{FD}}(E_{\bm{k}})\big\{1 - n_{\mathrm{FD}}
(E_{\bm{k}}) \big\}
\bigg(\frac{\partial \xi_{\bm{k}}}{\partial k_{\alpha}}\bigg)^{2} \Bigg]~. 
\nonumber
\end{eqnarray}

Our expressions (\ref{MFDeCorrFu-LOm}) - (\ref{MFPhasStif-LWa}) have been 
compared with Eqs.\! (16) - (19) of Ref.\! \onlinecite{Param-PRB2000}, taken 
in the low-frequency, long-wavelength limit we consider. We obtain the 
same results. Therefore we can conclude that $\mathcal{K}^{AA}$ and 
$\mathcal{K}^{\theta\theta}$ are equal, up to a constant factor proportional 
to the electronic charge. It provides an additional non-trivial check of 
Eq.\! (\ref{equal}).


\begin{thebibliography}{99}
\bibitem{Nagaosa_ref}
A detailed derivation of the superconductivity theory in the framework of 
path integral can be found in Chapter 5 of Ref. \onlinecite{Nagaosa_bookCM}.
\bibitem{DiagrNambu}
The diagrammatic rules in the framework of Nambu formalism are developed 
in Chapter 8 of Ref. \onlinecite{Rick-GreenF}.
\end{thebibliography}
\end{document}